\documentclass[aps,prc,twocolumn,superscriptaddress]{revtex4-2} 
\usepackage{amsmath,amssymb,amsthm} 
\usepackage[permil]{overpic} 
\usepackage[english]{babel}        

\begin{document}
\title{Generalized gauge-space rotations in atomic nuclei: A critical insight}

\author{Chong Qi}
\affiliation{Department of Physics, KTH Royal Institute of Technology, SE-10691 Stockholm, Sweden}

\author{Roberto J. Liotta}
\affiliation{Department of Physics, KTH Royal Institute of Technology, SE-10691 Stockholm, Sweden}

\author{Ramon Wyss}
\affiliation{Department of Physics, KTH Royal Institute of Technology, SE-10691 Stockholm, Sweden}

\date{\today}

\begin{abstract}
We critically reexamine the concepts of pairing rotations and moments of inertia in gauge space extracted from experimental binding energies. Our analysis focuses on pairing correlations among like nucleons, neutron-proton pairing, and $\alpha$-type correlations. By investigating $\alpha$ separation energies and binding-energy differences along chains of fixed isospin projection and subtracting macroscopic contributions, we reveal a remarkably smooth and nearly universal behavior in the residual $\alpha$ correlation energy. These results exhibit the parabolic trends characteristic of collective rotations in gauge space. We demonstrate that the standard definition of the gauge-space moment of inertia for like-nucleon pairing is dominated by macroscopic contributions from Coulomb and symmetry energies. Once these are removed, the remaining moment of inertia becomes negative. This suggests that the observed behavior reflects the loss of correlation energy due to Pauli-blocking effect. Our results indicate that $\alpha$ correlations constitute a genuine collective mode associated with quartetting dynamics arising from the coherent coupling of two superfluid components.

\end{abstract}

	\maketitle

A hallmark of quantum many-body physics is the emergence of coherent motion, whereby seemingly independent particles form collective states. Atomic nuclei exhibit three dominant modes of collectivity: vibration, rotation and pairing correlations. The interplay between pairing and rotations can be explored within the framework of generalized rotations associated with symmetry breaking. Quadrupole collectivity is manifested in rotational band structures characterized by an approximately constant moment of inertia in real space. In contrast, pairing collectivity can be described in terms of rotations in gauge space, with a moment of inertia associated with the pair density. This latter mode has been thoroughly investigated \cite{Bes1966,BM,Broglia1986-rh,Pot13} and has recently attracted renewed interest through approaches based on effective field theory \cite{Pap}, linear response theory \cite{PhysRevC.92.034321}, and extensions to a moment-of-inertia tensor in gauge space that incorporates neutron–neutron, proton–proton, and neutron–proton contributions \cite{Hinohara2016}.

In this letter, we critically examine the standard definition of the moment of inertia in gauge space, which includes a contribution from the liquid-drop energy that is not associated with pairing correlation. Upon removing
this macroscopic contribution, the resulting moment
of inertia changes amplitude and sign, revealing
the true contribution to the coherence of nucleonic motion at the Fermi surface. Beyond the two components of superfluidity associated with neutron and proton pairing, we demonstrate that the third component of rotations in gauge space must be built upon the collective dynamics of $\alpha$ correlation, which is governed by the coherent motion of proton and neutron pairs \cite{Qi2021-ak}.

We begin with a quantitative description of binding-energy differences in open-shell superfluid nuclei, starting from alpha-correlated states. Analogous to the definition of the two-nucleon separation energy, we introduce the $\alpha$ correlation energy induced by the addition of the last neutron and proton pairs as
\begin{equation}
E_{\alpha}(N, Z)=B(N, Z)-B(N-2, Z-2)-B_{\alpha}
\end{equation}
where $B$ stands for the (positive valued) binding energy. This definition is identical to that of a standard separation energy, except that the large intrinsic binding energy of the \(\alpha\) particle is subtracted. Consequently, the values of \( E_{\alpha} \) are uniformly shifted downward by a constant and thus quantify only the additional energy gain associated with incorporating the last proton and neutron pairs. With this convention, \( E_{\alpha} \) is simply the negative of the conventional \(\alpha\)-decay \(Q\) value.

\begin{figure*}[]
		\centering
		\begin{overpic}[width=0.48\textwidth]{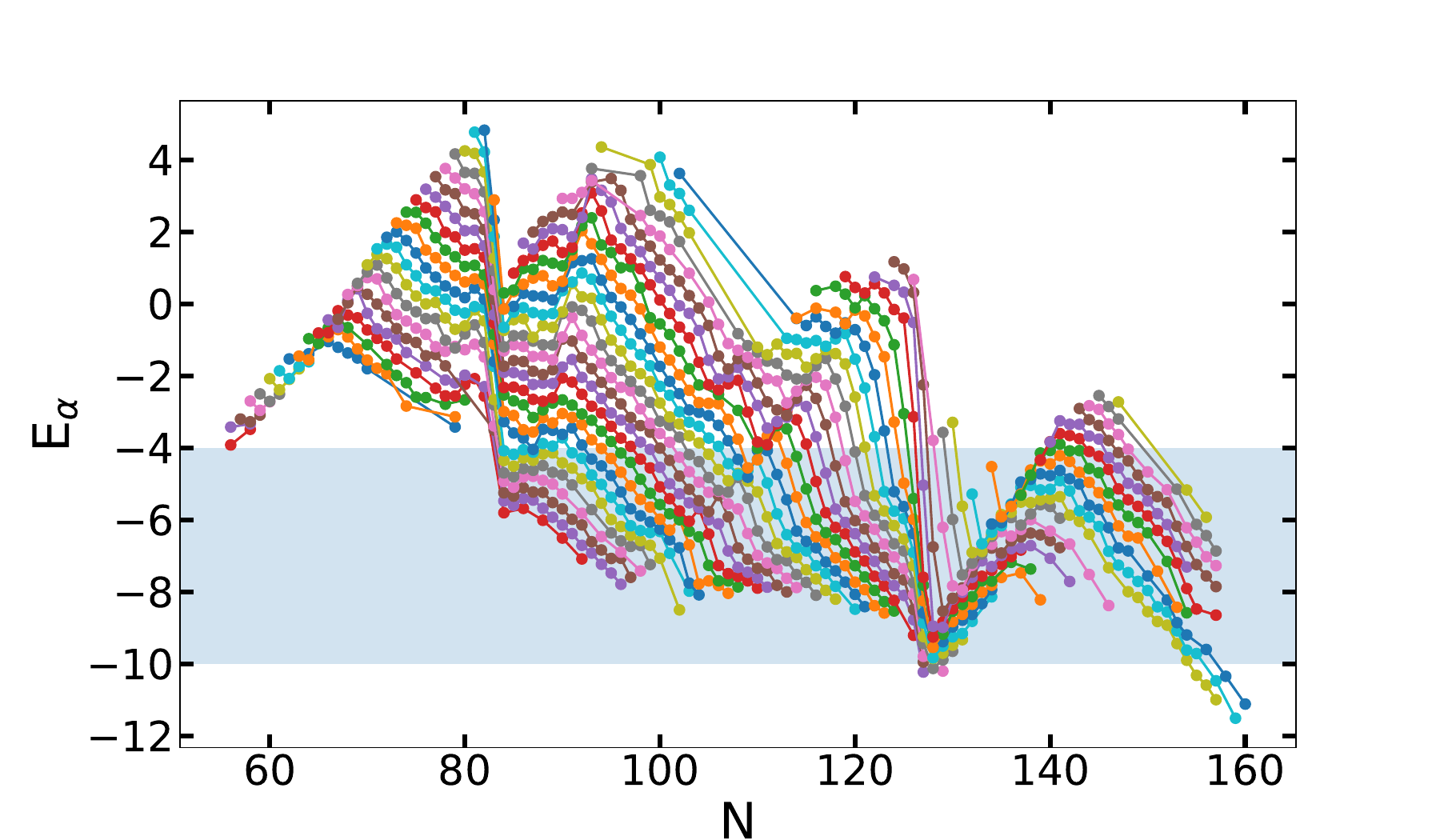}%
			\put(175,425) {(a)}%
		\end{overpic}
		\begin{overpic}[width=0.48\textwidth]{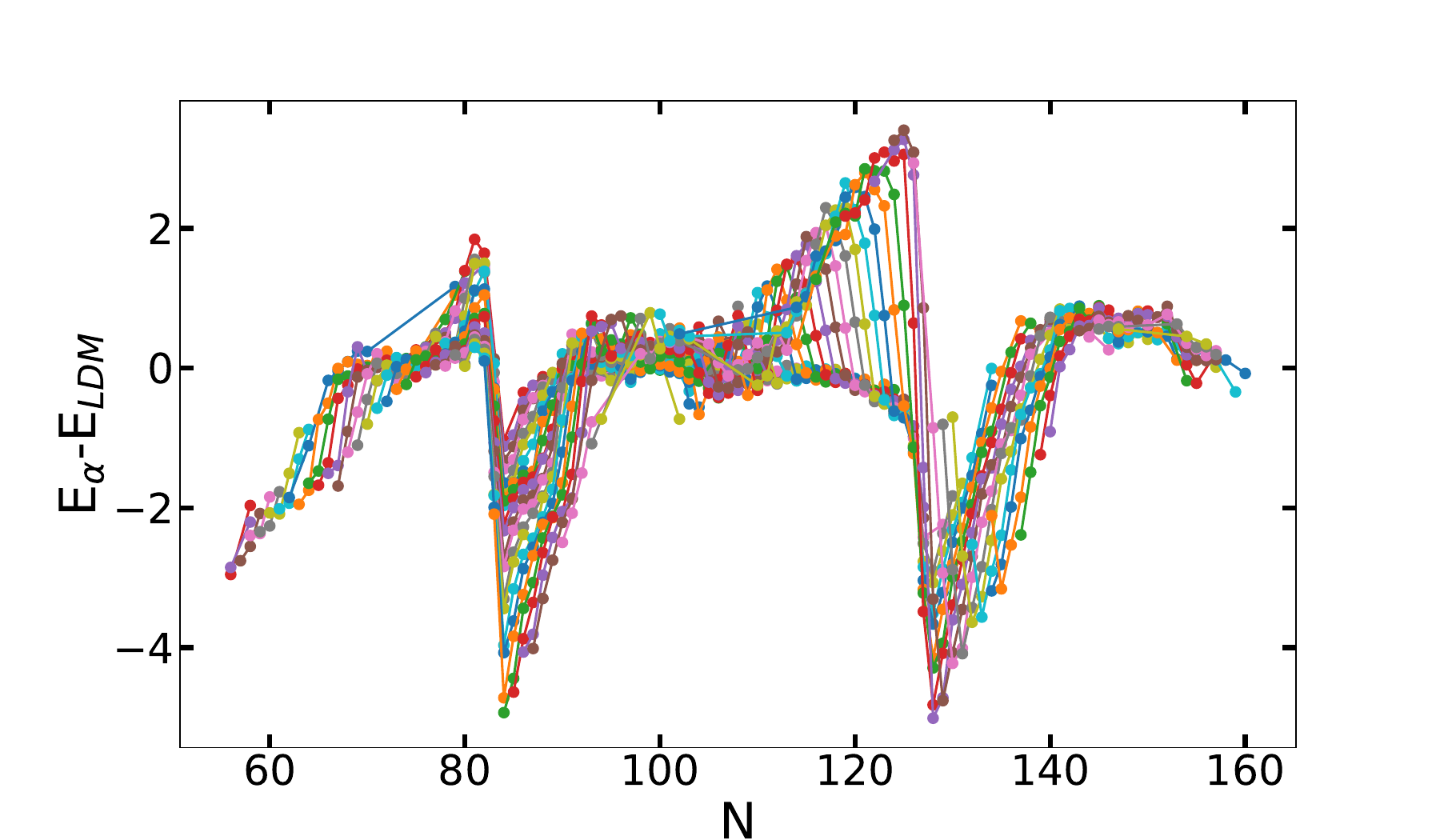}%
			\put(175,425) {(b)}%
		\end{overpic}
		\begin{overpic}[width=0.48\textwidth]{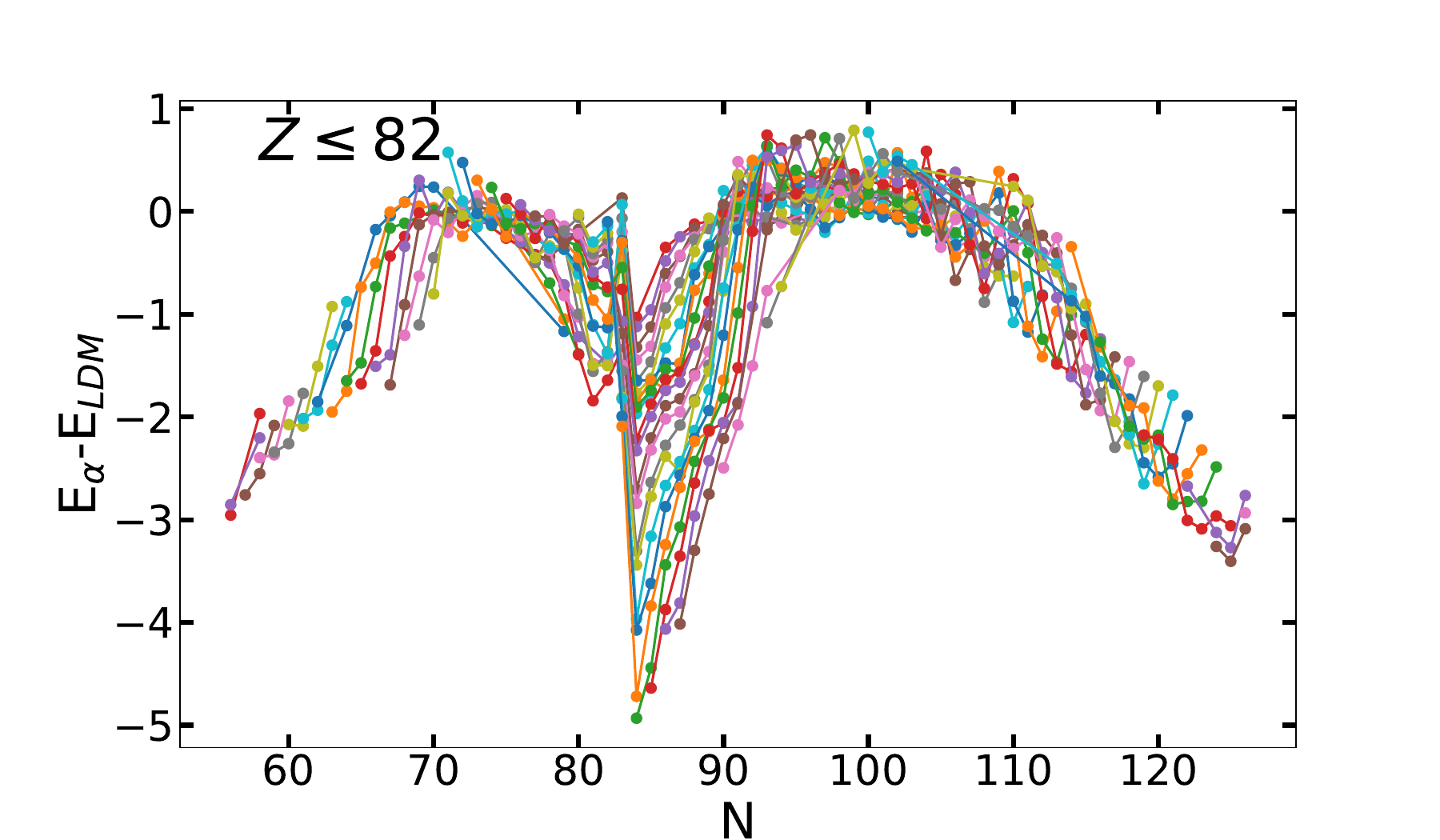}%
			\put(175,405) {(c)}%
		\end{overpic}
		\begin{overpic}[width=0.48\textwidth]{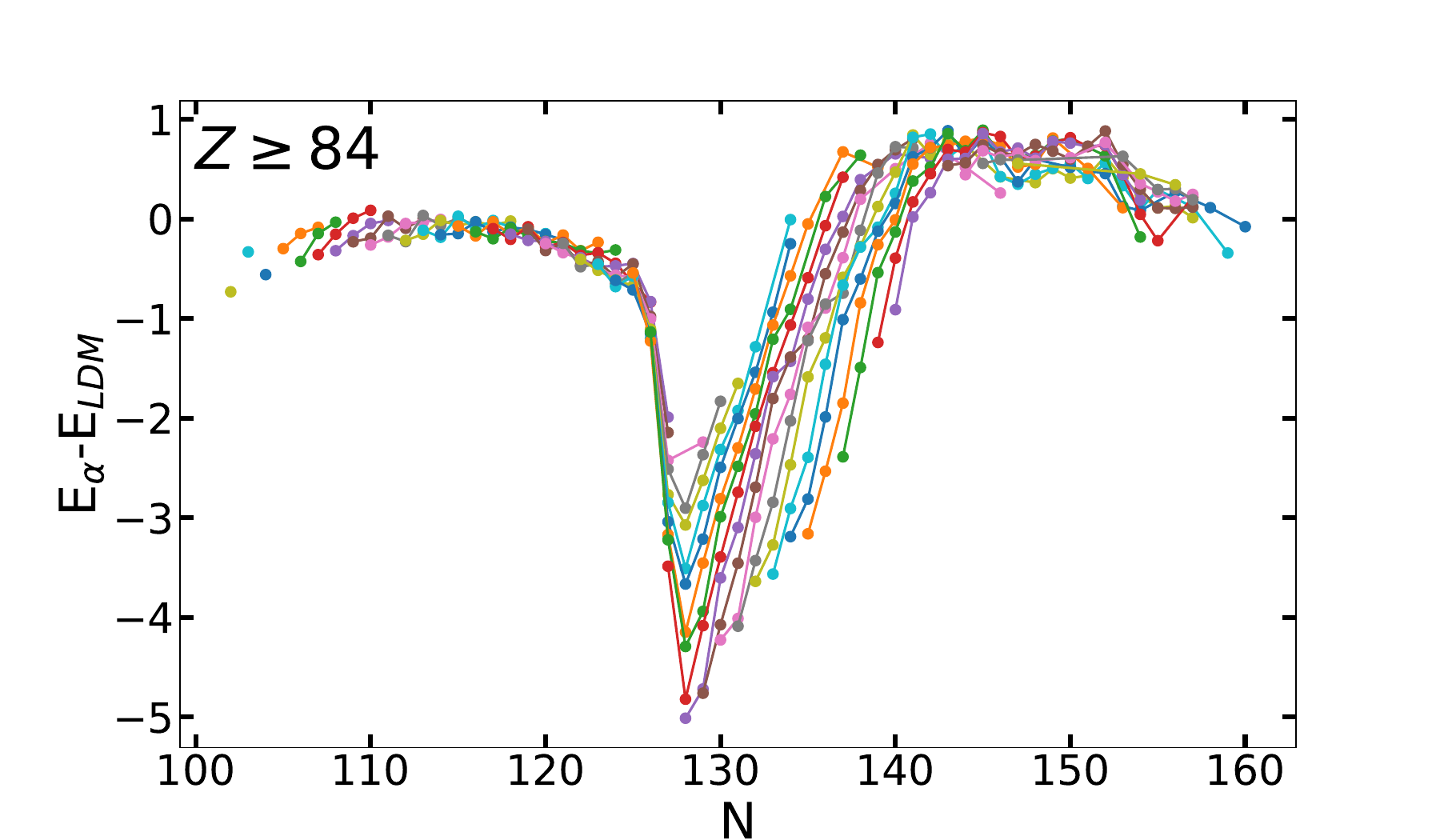}%
			\put(175,345) {(d)}%
		\end{overpic}
		\caption{Systematics of the $\alpha$-correlation energy $E_{\alpha}$ in the ground states of heavy nuclei as a function of neutron number, extracted from experimental atomic masses~\cite{wang2021ame}. Data points with the same isospin are connected by solid lines. (a) $E_{\alpha}$ obtained directly from experimental binding energies. (b) $E_{\alpha}$ after subtracting Coulomb and liquid-drop contributions from the total binding energy. (c) Same as panel (b), but shown for $Z\leq82$ relative to the mid-shell, taking particle--hole conjugation into account. (d) Same as panel (c), but for $Z\geq84$.}
		\label{dssd}
	\end{figure*}

The \( E_{\alpha} \) values extracted from experimental binding energies are shown in the upper-left panel of Fig.~\ref{dssd}a for nuclei above \(^{100}\)Sn. In the figure, nuclei with the same isospin projection \( T_z \) (or, equivalently, belonging to the same \(\alpha\)-decay chain) are connected by lines. All nuclei, namely even-even, odd-\(A\), and odd-odd systems, are treated on an equal footing.
Most observed \(\alpha\) emitters exhibit \( E_{\alpha} \) values in the range of approximately \(-4\) to \(-10\) MeV, indicated by the gray-shaded region in the figure. This is unsurprising, since this energy window corresponds to the regime in which \(\alpha\) decay is energetically favored. However, it also leads to an intriguing conclusion: nuclei that readily undergo \(\alpha\) decay actually possess the smallest \(\alpha\)-correlation energy. Furthermore, \(E_{\alpha} \) displays an overall decreasing trend with increasing mass number, reflecting the increasingly negative contribution of the Coulomb interaction in heavy nuclei.

To isolate the impact of the nuclear interaction between valence nucleons on the \(\alpha\)-correlation energy, we subtract the bulk contributions to the binding energy (denoted as $B_M$ in Eq.~(\ref{dssd-m}) below) and thus to \(E_{\alpha}\) arising from smooth macroscopic effects. These include the Coulomb energy as well as the macroscopic terms of the liquid-drop model. The resulting quantity,
\begin{equation}\label{dssd-m}
\begin{aligned}
E'_{\alpha}(N, Z) ={}& B(N, Z) - B_{M}(N, Z) \\
& - B(N-2, Z-2) + B_{M}(N-2, Z-2) - B_{\alpha}.
\end{aligned}
\end{equation}
is shown in Fig.~\ref{dssd}b.

A striking feature of this figure is that all curves display a regular behavior, with pronounced minima near the magic numbers. As expected, the overall decreasing trend disappears once the Coulomb contribution is removed. Even more remarkably, the curves cluster closely together and exhibit nearly identical evolution. This indicates that the large spreading seen in the experimental data of Fig.~\ref{dssd}a is dominated by macroscopic contributions, primarily from the Coulomb and symmetry energies.
The positive contribution from the symmetry energy, which scales approximately as \(T(T+1)/A\), increases with the isospin quantum number \(T\), while the negative contribution from the Coulomb energy, proportional to \((A/2 - T)^2 / A^{1/3}\), becomes less negative as \(T\) increases.

Some of the curves exhibit a noticeably asymmetric behavior relative to the mid-shell, decreasing as one approaches the lower shell closure but increasing toward the upper one. This asymmetry arises because, with $E_{\alpha}$ as defined above, one measures particle correlations relative to a core with two pairs removed. However, hole correlations dominate when approaching a shell closure from below.

To account for this, in Fig.~\ref{dssd}c and \ref{dssd}d we first separate nuclei below and above $Z=82$ and, to preserve particle–hole symmetry, introduce a minus sign when approaching the neutron shell closures $N=82$ or $126$ from below. For nuclei below the shell closures, this gives
\begin{equation}
\begin{aligned}
E'_{\alpha}(N, Z) ={}& -B(N, Z) + B_{M}(N, Z) \\
& + B(N-2, Z-2) - B_{M}(N-2, Z-2) - B_{\alpha},
\end{aligned}
\end{equation}
resulting in curves that display an almost parabolic behavior.  In  Fig.~\ref{dssd}b-d, only relative variations of $E_{\alpha}$ are physically meaningful, since the absolute values can depend on the specific prescription used to remove macroscopic (Coulomb and liquid-drop) contributions.
It is important to emphasize that we are primarily interested in this parabolic trend of the relative values, which signals the change in the $\alpha$ correlation energy, rather than the absolute $E_{\alpha}$ values. 

This simple analysis reveals that the $\alpha$ correlation energy in atomic nuclei exhibits a regular and collective pattern, depending primarily on the number of valence $\alpha$ particles (or holes). It corroborates our previous work on $\alpha$ decay, showing that
$\alpha$ decay is a collective phenomenon \cite{Qi2021-ak}, that may be compared to like-particle pairing correlations as a general collective phenomenon that appear across quantum many-body systems. In atomic nuclei, pairing correlations are often described by the BCS approach, that breaks particle
number. The symmetry breaking of the BCS solution can be described in terms of
pairing rotations in gauge space~\cite{Bes1966,Bayman1969,Brink2005,Broglia2000,Clark2006,Potel2011}.

		\begin{figure*}[!htb]
		\centering
		\begin{overpic}[width=0.38\textwidth]{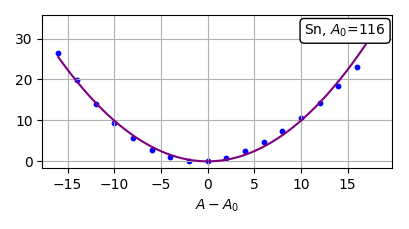}%
		\end{overpic}
		\begin{overpic}[width=0.38\textwidth]{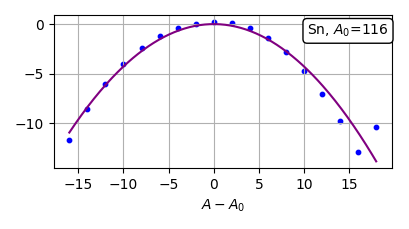}%
		\end{overpic}
		\begin{overpic}[width=0.38\textwidth]{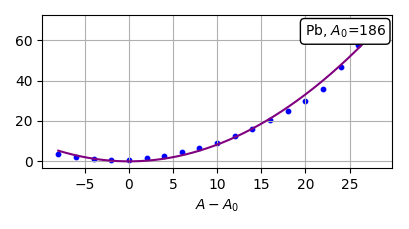}%
  \put(-20,20){\rotatebox{90}{Quadratic Residual}}
  \end{overpic}
		\begin{overpic}[width=0.38\textwidth]{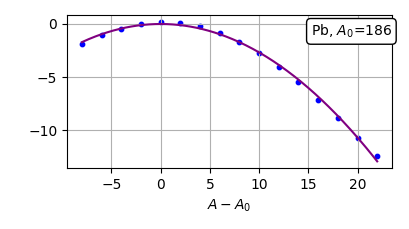}%
		\end{overpic}
        		\begin{overpic}[width=0.38\textwidth]{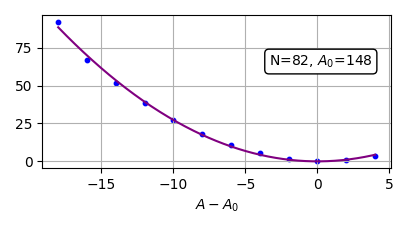}%
		\end{overpic}
		\begin{overpic}[width=0.38\textwidth]{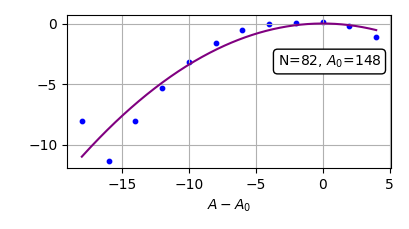}%
		\end{overpic}
		\caption{Left: Quadratic residual energy (in MeV) extracted from experimental binding energies (solid symbol) for even-even Sn and Pb isotopes and $N=82$ isotones. Right: Same as left but with the macroscopic symmetry energy and Coulomb energy subtracted from the binding energy. We adopt the sign convention used in Refs.~\cite{Pap,Hinohara2016}. Because of this choice, the trends shown here are the inverse of those in Fig.~\ref{dssd}: a concave-upward curve in the left panels indicates enhanced extra binding energy for nuclei near the center value $A_0$. In contrast, concave-downward curves (right) show enhanced binding energy for systems with fewer pairs as they move away from $A_0$. }
		\label{pair-rotate}
	\end{figure*}

The pairing rotational energy for a system with mass number $A$ can be expressed as a quadratic function \cite{Bayman1969,Brink2005,Krappe1975,Beck1972,Hinohara2016,Pap}:
\begin{equation} \label{quad}
E(A) = E(A_0) + \lambda_{A_0} \, (A - A_0) + \frac{(A - A_0)^2}{2 \mathcal{J}_{A_0}},
\end{equation}
where $E(A)$ is the \emph{negative} total energy of the system $A$, $A_0$ denotes a chosen reference nucleus and can be interpreted as the average particle number of the ground-state configuration associated with the pairing rotation. The variable $A$ runs over the nuclei in a given isotopic or isotonic chain. The quantity $\mathcal{J}_{A_0}$ is the moment of inertia in gauge space, which is related to the pairing density and depends on the choice of $A_0$, while $\lambda_{A_0}$ is the chemical potential. The three coefficients in Eq.~\eqref{quad} have historically
be determined by fitting to the measured binding energies
of the chosen set of nuclei.
It is not  obvious that the moment of inertia in gauge space can be determined by a
fit to differences in binding energies.

Typical examples that have been intensively studied include the long semi-magic Sn and Pb isotopic chains, as well as the $N=82$ isotonic chain, as shown in Fig.~\ref{pair-rotate} (see also Fig.~1 in Ref.~\cite{Potel2011}, Fig.~9 in Ref.~\cite{Pot13}, Fig.~1 in Ref.~\cite{Hinohara2016} and Figs.~2–4 in Ref.~\cite{Pap}). For Sn isotopes between $N=50$ and 82, a nearly symmetric behavior is observed when $A_0=116$ ($N_0=66$), corresponding to the exact mid-shell. The system loses energy as it deviates from this mid-shell configuration. In practice, the choice of $A_0$ is arbitrary, but it does not affect the characteristic quadratic shape of the energy curve.

Although one often links pairing correlations to the quadratic nature of pairing rotational energy, experimental binding energies do not provide direct evidence for this. Instead, macroscopic symmetry and Coulomb energies impose a strong quadratic dependence on the mass number within isotopic and isotonic chains. As a result, one must disentangle these macroscopic contributions before attributing the observed quadratic behavior to pairing correlations.
For that, we investigate how the quadratic behavior changes when the symmetry and Coulomb energy contributions are subtracted from the experimental binding energies. This procedure is illustrated in the right panel of Fig.~\ref{pair-rotate} for the same three nuclear chains. Once these macroscopic contributions are removed, the quadratic dependence is significantly reduced. Even more strikingly, the curvature of the parabola changes sign, demonstrating that the dominant quadratic trend in the raw binding energies originates from macroscopic effects rather than from pairing correlations. Our analysis results in a deeper understanding
of the valence space dependence of pairing correlations.

This observation highlights the importance of interpreting with care the coefficients extracted from Eq.~(\ref{quad}) and underscores the necessity of removing macroscopic contributions when attempting to quantify genuine pairing effects.
Nevertheless, one may still ask to what extent pairing correlations themselves contribute to the residual quadratic dependence. Insight into this question can be obtained from the exact or quasi-exact solution of the pairing Hamiltonian, or from approximate approaches such as the BCS method (See Appendix). For an even–even system containing $n$ pairs, the exact ground-state energy $E$ (relative to a presumed core) and pair separation energy $S_2$ can be well approximated as \cite{Changizi2015}
\begin{eqnarray}\label{exact-energy}
    E(n) \simeq  &~~ nE_2 + n(n-1)\,\mathcal{G},\\
S_{2}(n) \simeq &  -E_2-2(n-1)\mathcal{G}
\end{eqnarray}
where $E_2$ denotes the (negative) ground-state energy of a system containing a single pair. One immediately notes that pairing correlations contribute directly to the linear term in $n$. The constant $\mathcal{G}$ is \textit{positive} and can be determined from the exact solutions of the pairing Hamiltonian. The above expression becomes exact for systems confined to a single-$j$ shell where $\mathcal{G}$ reduces to the pairing strength $G$. The quadratic term proportional to $n(n-1)$ reflects the effect of the Pauli principle:
as the number of pairs increases, the phase space available for forming additional pair correlations is progressively
reduced, leading to a \textit{loss} of correlation energy per added pair. 
That is the reason why the moment of inertia term $\mathcal{J}_{A_0}$ in Eq.~(\ref{quad}) changes sign
when the liquid-drop contribution is removed from experimental masses, as illustrated in the right panel of Fig.~\ref{pair-rotate}. As a result, the two-particle (pair) separation energy \textit{decreases} as the number of pairs increases.

The apparently positive value of $\mathcal{J}_{A_0}$ in the left panels of Fig. \ref{pair-rotate} in fact corresponds to a negative microscopic moment of inertia, which produces additional binding for nuclei near the chosen reference point. The problem originates from the choice of reference point, which is typically taken to be mid-shell and leads to ambiguities in counting the number of pairs and, consequently, the actual direction, increase or decrease, of the correlation energy. This behavior contrasts with what is expected from purely fermionic systems, where a binding-energy curve driven solely by like-nucleon pairing bends downward and is associated with a positive moment of inertia. That positive value of $\mathcal{J}$ extracted from fits to binding-energy data was historically misinterpreted as a positive pairing moment of inertia, a misunderstanding that has contributed to confusion in the literature for several decades.
The change of sign of the moment of inertia $J_{A_0}$, extracted from experimental binding
energies after subtraction of macroscopic contribution, can be readily understood from the expression for the moment of inertia in gauge space. When the symmetry energy dominates the definition of 
$\lambda$, i.e., when the difference between the two-neutron separation energies is large, it produces a positive contribution. However, once the symmetry-energy contribution is removed from the two-neutron separation energy, the term changes sign and exhibits a completely different curvature.
The effect
should be understood as a manifestation of this exact fermionic physics and the reduced incremental gain in
correlation energy imposed by the Pauli principle as additional pairs are added to the system.

It should be emphasized that, as can be seen from Fig.~\ref{pair-rotate}, the smoothness of the parabola is preserved even after the subtraction of macroscopic contributions. That is a consequence of the pairing mode. In contrast, an uncorrelated system would exhibit discontinuities whenever a new set of single-particle orbitals becomes occupied. This persistence of a smooth quadratic trend supports the notion of rotations in gauge space with an approximately constant moment of inertia, related to the coherent nature of pair transfer.
For protons, an analogous behavior is observed: the moment of inertia changes after subtracting the Coulomb and symmetry-energy contributions, while the overall regularity of the curve remains comparable to that of neutrons. 

This naturally raises the question of whether the concept of pairing correlations can be extended to neutron–proton correlations.
Within this context, the neutron–proton pairing rotational energy has been introduced as \cite{Krappe1975,Hinohara2016}
\begin{equation} \label{quadnp}
E_{np}(A)=\frac{(Z-Z_0)(N-N_0)}{2\mathcal{J}_{np}}.
\end{equation}
The associated neutron–proton moment of inertia was extracted from the double difference of binding energies,
\begin{equation}
\begin{aligned}
\frac{1}{\mathcal{J}_{np}}
&= \delta V_{pn}(Z,N)
= \frac{1}{4}\Big[B(Z,N)-B(Z,N-2) \\
& \quad - B(Z-2,N)+B(Z-2,N-2)\Big] \\
&\approx \frac{2a_{\rm sym}}{A},
\end{aligned}
\end{equation}
This method of extracting the average proton--neutron interaction was proposed and explored in a series of works by Casten \textit{et al.}~\cite{Zhang1989-hh,PhysRevLett.94.092501} and has long been regarded as a probe of neutron--proton correlations (see, for example, Refs.~\cite{Zhang21,80Zr,Wang23,Kaneko2003}). However, it is important to recognize that the bulk behavior of this quantity extracted from experimental binding energies is, again, dominated by the symmetry-energy contribution \cite{Qi2012}. This follows directly from the form of the symmetry energy, which is proportional to $a_{\rm sym}(N-Z)^2/A$, together with the identity
\[
(N-Z)^2 + (N-2 - Z + 2)^2 - (N - Z - 2)^2 - (N - Z + 2)^2 = -8,
\]
valid for $|N-Z| \geq 2$.
After removing the symmetry energy from the double differences,
there are no average correlation left that can provide a regular moment of inertia.

		\begin{figure}[!htb]
		\centering
		\begin{overpic}[width=0.49\textwidth]{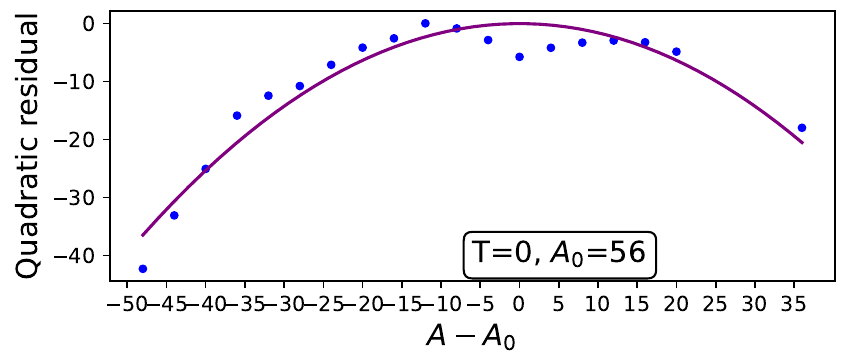}%
		\end{overpic}
		\begin{overpic}[width=0.49\textwidth]{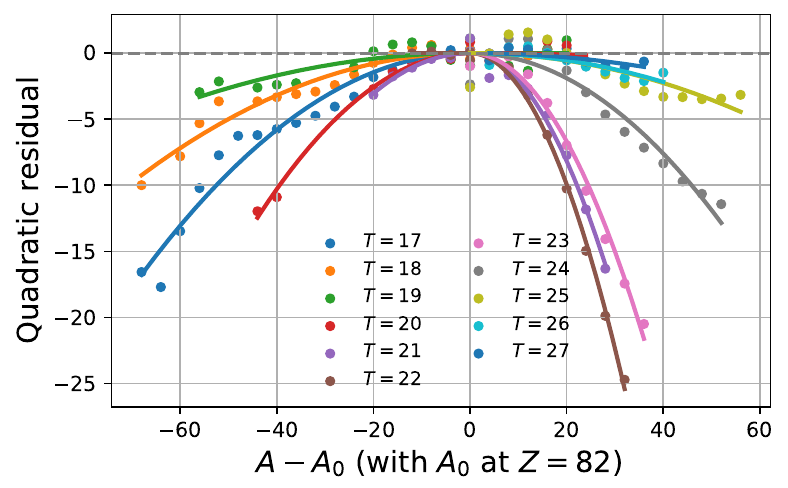}%
		\end{overpic}
		\caption{Quadratic residuals (solid symbols) extracted from experimental binding energies after subtracting the Coulomb contribution for various $\alpha$ chains with fixed isospin projection. Upper panel: The $T=0$ chain is plotted relative to $^{56}$Ni. Lower panel: Heavy isobar chains are plotted relative to their corresponding Pb isotopes. The gain in extra quadratic energy as $\alpha$ particles or holes are added to the Pb core is consistent with the $E_{\alpha}$ systematics shown in Figs. \ref{dssd}c and \ref{dssd}d.}
		\label{alpha-rotate}
	\end{figure}

A more proper degree of freedom to study the pairing rotational energy is along chains with ﬁxed isospin projection values (for example, a $\alpha$-decay chain). By defining the single particle amplitude in $\alpha$ decay, our recent works have revealed that the amplitude of $\alpha$ decay can only be accounted for by the presence of strong $\alpha$ correlations \cite{clw,Qi2021-ak}. More generally, related approaches emphasize quartetting as a manifestation of $\alpha$-like four-body correlations in nuclei \cite{Sandulescu2012,PhysRevC.102.061301,Baran2020QuartetBCS}.  Hence, one expects the appearance of a corresponding smooth moment of inertia in gauge space reflecting the collective pairing modes in atomic nuclei corresponding to like-nucleon, neutron-proton and $\alpha$ or quartet like correlations. 
The study of correlations within the concept of gauge rotations,  as induced by the addition or removal of an alpha particle, was performed in Ref.~\cite{Pap} for a chain centered with $^{166}$Y. Indeed a rather symmetric parabolic behavior was observed. For differences in binding energies along $\alpha$ chains, the macroscopic symmetry energy effect does not contribute and only the change in Coulomb energy does. 

In Fig. \ref{alpha-rotate} we plotted the quadratic residual of the pairing rotational energy along the long $N=Z$ chain and $\alpha$ chains of heavy nuclei around Pb isotopes.
The smooth quadratic behavior as seen in the figure is a result of the smooth $\alpha$ energy systematics as presented in Fig. \ref{dssd}.
Since lines of constant $T_z$ cross shell gaps, the curves exhibit some shell structure and are less smooth than the corresponding curves for neutrons and protons. This is
evident in particular along the 
$N=Z$ line, where $^{56}$Ni is chosen as
a reference nucleus and exhibits a drop. In spite of the visibility of shell gaps, the curve has a distinct parabolic behavior, pointing to the smoothness of 
$\alpha$ correlations. The same trend is observed for nuclei around the Pb isotopes, indicating that 
$\alpha$ correlations remain smooth even near heavier shell closures, which is manifested from systematics of $\alpha$ decay formation probabilities.

We can model the nuclear system as $n$ interacting identical $\alpha$ particles, with energy  $\lambda_\alpha$, treated as structureless quasi bosons outside a certain core.
Let's assume the correlation is induced by $\alpha$ particles interacting through a constant monopole pairing interaction of strength $G$ which gives a simple Hamiltonian  of the form
\[
H = \lambda_\alpha \hat n - G P^\dagger P.
\]
where $\hat n$ is the $\alpha$-particle number operator and $P^\dagger$ creates an $\alpha$ pair.
The total energy of the system is given by
\begin{eqnarray}\label{ealpha}
    E(N) = n \lambda_\alpha - n(n-1)\frac{G}{2},\nonumber\\
    =n(\lambda_\alpha+\frac{G}{2})-\frac{n^2}{2\mathcal{J}_{\alpha}}
\end{eqnarray}
where the second term accounts for the \emph{attractive} correlation between all distinct $\alpha$ pairs. The total energy of the system can then be parametrized in a  way similar to Eq. (\ref{quad}):
\begin{equation} \label{quad-alpha}
E(A) = E(A_0) + \lambda_{\alpha,A_0} \, (A - A_0) - \frac{1}{16}\frac{(A - A_0)^2}{2 \mathcal{J}_{\alpha,A_0}},
\end{equation}
where the factor of $1/16$ is introduced to ensure consistency with the previous definition, which assumes the propagation of one $\alpha$ particle ($\Delta A = 4$) at each step.
The most striking difference between the $\alpha$ correlation energy and the pairing correlation described by Eq.~(\ref{quad}) is the opposite sign of the quadratic term. This difference reflects the fermionic nature of nucleon pairing, which is governed by Pauli blocking, in contrast to the quasi bosonic character of $\alpha$ particles, for which correlations accumulate coherently. This explains the parabolic behavior of the $\alpha$ correlation energy observed in Fig. \ref{dssd} as well as the quadratic behavior of the binding energy as seen in Fig. \ref{alpha-rotate}. One can expect the $\alpha$-correlation energy to saturate at mid-shell, where the number of effective $\alpha$ particle/hole configurations is maximized.

One may therefore argue that the expression of $\alpha$ rotation in gauge space, applied to $\alpha$-chains of fixed isospin
projection, provides a suitable framework for simultaneously studying like-particle pairing and possible neutron--proton pairing effects as well as a more natural framework for studying pairing rotational energies extracted from binding energies. An important feature of this approach is that it is free from macroscopic symmetry-energy contributions and, at the same time, offers a direct measure of quartetting correlations induced by the coherent coupling of collective neutron and proton components, highlighting its composite and collective character.

From a general perspective, nuclear binding energies and spectra have long been interpreted in terms of elementary collective modes, most prominently pairing correlations among like nucleons \cite{BM,BMP,DH} and the formation of clusters, in particular $\alpha$ particles \cite{Duss82,Duss88}. Substantial $\alpha$ correlations are well established in excited states near the $\alpha$-decay threshold, such as the Hoyle state, and have been extensively investigated both experimentally and theoretically \cite{Freer14,Ots22,Ikeda68,Freer18}. At the same time, $\alpha$ decay itself provides direct evidence for $\alpha$ clustering already present in the ground states of heavy nuclei. Complementary information has recently become accessible through $\alpha$ knockout reactions, which are emerging as a sensitive probe of $\alpha$ clustering in $\alpha$-bound systems \cite{RC77,TY21,Tan21,Yos18}. Microscopically, $\alpha$ clustering is inhibited in the dense nuclear interior by the Pauli exclusion principle, yet it emerges naturally at the nuclear surface, where it is further enhanced by nucleon pairing correlations \cite{clw}.
The systematics of $\alpha$-cluster formation probabilities show an increasing trend with the number of $\alpha$ particles as the system deviates from major shell closures, followed by saturation at mid-shell. This behavior is consistent with the $\alpha$-correlation energy presented in this work.

To summarize, we demonstrated that $\alpha$ correlations exhibit a nearly universal, smooth behavior across the nuclear chart. These correlations display parabolic trends characteristic of rotations in gauge space, persisting despite shell effects. This indicates that $\alpha$ correlations constitute a collective mode distinct from conventional pairing, emerging from the coherent coupling of two superfluid components—a signature of quartetting dynamics rather than simple neutron-proton pairing. 

Standard moments of inertia in gauge space extracted from experimental binding energies are dominated by macroscopic liquid-drop terms, specifically symmetry and Coulomb energies, which obscure the underlying pairing dynamics. Removing these macroscopic components fundamentally redefines the resulting moments of inertia. This clarifies why neutron-proton double binding-energy difference is not a clean probe of collective neutron-proton pairing: their bulk behavior is governed by the symmetry energy,  leaving little trace of a collective
neutron-proton pairing rotational mode.

Along the $\alpha$ chains (fixed isospin projection), the symmetry energy remains constant. Consequently, $\alpha$ rotations in gauge space for $\alpha$ chains provide provide a unified framework for simultaneously probing like-particle pairing, neutron–proton correlations, and genuine quartetting. In this context, one expects correlations to be enhanced as the number of $\alpha$ particles increases. The observed enhancement of $\alpha$-decay strength, together with the remarkably smooth systematics of $\alpha$ correlation energies, points to $\alpha$ clustering as a fundamental collective mode of the nuclear many-body system. This perspective is essential for interpreting the so-called “superallowed” $\alpha$ decay often anticipated near $^{100}$Sn and commonly attributed to neutron–proton pairing. Our analysis instead suggests that $\alpha$ correlations may be significantly enhanced in mid-shell nuclei due to the coexistence and mutual coupling of multiple $\alpha$ particles, giving rise to a collective background that can obscure effects associated with simpler systems, such as $^{104}$Te, where only a single $\alpha$ degree of freedom is active.

\vspace{1cm} 
\section*{ACKNOWLEDGMENTS}
We thank the financial support from the Olle Engkvist Foundation and the computational resources provided by the National Academic
Infrastructure for Supercomputing in Sweden (NAISS) at PDC, KTH.

\appendix

\section{Gauge symmetry breaking and the pairing rotational Mode: from BCS to number-conserving pairing}
\label{app:pairing_rotation}
The purpose of this appendix is to clarify the physical meaning of
\emph{pairing rotation} and its relation to the particle-number dependence
of the ground-state energy.
The particle-number operator $\hat N$ can be associated with a gauge
(phase) transformation in an abstract gauge space. The corresponding
gauge angle $\phi$ is canonically conjugate to the particle number and
satisfies the commutation relation
\begin{equation}
[\phi,\hat N] = i .
\end{equation}
In the number representation, this implies
\begin{equation}
\phi = i \frac{\partial}{\partial N}.
\end{equation}
A global gauge transformation generates a continuous set of states,
\begin{equation}
|\Phi(\phi)\rangle = e^{i\phi \hat N} |\Phi(0)\rangle,
\end{equation}
labeled by the gauge angle $\phi$, where $\hat N$ is the particle-number
operator. Within the mean-field approximation, all states related by such
a transformation are \emph{degenerate in energy}.

Historically, the operator $\hat N$ was interpreted by analogy with
angular momentum, and gauge transformations were described as rotations
in gauge space associated with changes in particle number. However, this
terminology is somewhat confusing: the transformation is not a spatial
rotation but a global phase transformation.

In mean-field descriptions of pairing correlations, such as BCS or HFB
theory, particle-number conservation is realized through the introduction
of a Lagrange multiplier $\lambda$. The Hamiltonian then takes the form
\begin{equation}
H = H_0 - \lambda \hat N ,
\label{eq:H_lambda}
\end{equation}
where $H_0$ contains the kinetic energy and pairing interaction, and
$\lambda$ fixes the average particle number, playing the role of a
chemical potential.

The time evolution of the gauge angle follows from the Heisenberg
equation of motion,
\begin{equation}
\dot{\phi}
= \frac{i}{\hbar} [H,\phi]
= \frac{1}{\hbar} \frac{\partial H}{\partial N}
= \frac{\lambda}{\hbar}.
\label{eq:phi_dot}
\end{equation}
One gets,
\begin{equation}
\frac{\partial \dot{\phi}}{\partial N}
= \frac{1}{\hbar} \frac{\partial \lambda}{\partial N}
= \frac{1}{\hbar^{2}} \frac{\partial^{2} E}{\partial N^{2}} .
\end{equation}
This relation defines the \emph{pairing moment of inertia} $\mathcal{J}$,
\begin{equation}
\mathcal{J}^{-1}
= \frac{\partial^{2} E}{\partial N^{2}}
= \frac{\partial \lambda}{\partial N}.
\label{eq:J_pair}
\end{equation}
If the pairing moment of inertia is approximately constant, the energies
of neighboring even nuclei can be expanded in a quadratic form with
respect to particle number, as in Eq.~(\ref{pair-rotate}). This quadratic
dependence corresponds to a \emph{pairing rotational band} in
particle-number space. Indeed, the isotopic/isotonic chains shown in the left panel
of Fig.~\ref{pair-rotate} are commonly presented as illustrative examples
of pairing rotation.

It should be emphasized, however, that the gauge transformation defined
above does not by itself guarantee the emergence of a quadratic energy
dependence or a nonzero pairing moment of inertia. The appearance of a
pairing rotational pattern depends on the specific properties of the
mean-field Hamiltonian and the underlying pairing correlations. If the chemical potential is influenced by mean-field effects beyond pairing, as illustrated in the left panel of Fig.~\ref{pair-rotate}, the simple quadratic approximation of the ground-state energy no longer represents pure pairing correlations.

\subsection{Analytical BCS solution in a single-$j$ shell}

We consider a single degenerate shell with total degeneracy
\(2\Omega = 2j+1\), occupied by \(N\) particles interacting via a
constant pairing force of strength \(G\).
Owing to the degeneracy of the shell, the BCS amplitudes are 
\begin{equation}
N = 2\Omega v^2 ,
\qquad
v^2 = \frac{N}{2\Omega},
\qquad
u^2 = 1 - \frac{N}{2\Omega}.
\label{eq:v2_single_j}
\end{equation}
The gap equation yields
\begin{equation}
\Delta
= G \sum_{k>0} u v
= G \Omega u v
= G \Omega
\sqrt{\frac{N}{2\Omega}\left(1-\frac{N}{2\Omega}\right)} .
\label{eq:gap_single_j}
\end{equation}

The BCS ground-state energy of the degenerate shell is
\begin{equation}
E(N)
= 2\epsilon \sum_{k>0} v^2
- G \left(\sum_{k>0} u v \right)^2 ,
\end{equation}
which
can be rewritten as
\begin{equation}
E(N)
= \epsilon N
- \frac{G}{4} N (2\Omega - N).
\label{eq:E_single_j}
\end{equation}
Here $\epsilon$ denotes the single-particle energy.
The BCS approximation approaches the exact solution in the large $\Omega$ limit and exhibits a repulsive quadratic dependence on particle
number.

The chemical potential is obtained as the derivative of the energy with
respect to particle number,
\begin{equation}
\lambda(N)
= \frac{\partial E}{\partial N}
= \epsilon + \frac{G}{2}(N - \Omega).
\label{eq:lambda_single_j}
\end{equation}
The chemical potential thus varies linearly with the filling of the
shell and equals the single-particle energy \(\epsilon\) at mid-shell
(\(N=\Omega\)), reflecting particle–hole symmetry.

The curvature of the energy defines the pairing rotational inertia,
\begin{equation}
\frac{\partial^2 E}{\partial N^2}
= \frac{d\lambda}{dN}
= \frac{G}{2},
\end{equation}
which implies
\begin{equation}
\mathcal{J}_{\mathrm{single}}
= \left(\frac{\partial^2 E}{\partial N^2}\right)^{-1}
= \frac{2}{G}.
\label{eq:J_single_j}
\end{equation}

The single-$j$ shell provides an ideal realization of pairing, where the quadratic dependence of the ground-state energy on particle number is exact and the pairing moment of inertia $\mathcal{J}$ remains constant, independent of the filling. This behavior is consistent with our analysis shown in the right panel of Fig.~\ref{pair-rotate}, after subtracting the symmetry energy, and agrees with results from exact pairing solutions. It offers a simple microscopic illustration of pairing rotation as a collective motion in gauge space and serves as a useful benchmark for more realistic, multi-shell systems.

\subsection{General multi-$j$ BCS}

For a general set of doubly-degenerate single-particle levels $\epsilon_k$, the particle-number equation reads
\begin{equation}
    N = \sum_k 2v_k^2 = \sum_k \left( 1 - \frac{\epsilon_k - \lambda}{E_k} \right),
\end{equation}
where
\[
    E_k = \sqrt{(\epsilon_k - \lambda)^2 + \Delta^2}.
\]

Assuming that the pairing gap $\Delta$ is constant, we have
\begin{align}
    \frac{\partial v_k^2}{\partial \lambda} 
    &= -\frac{1}{2} \frac{\partial}{\partial \lambda} \left( \frac{\epsilon_k - \lambda}{E_k} \right) \nonumber\\
    &= -\frac{1}{2} \frac{-E_k - (\epsilon_k - \lambda) \frac{\partial E_k}{\partial \lambda}}{E_k^2} \nonumber\\
    &= \frac{1}{2E_k^2} \left( E_k - \frac{(\epsilon_k - \lambda)^2}{E_k} \right) \nonumber\\
    &= \frac{\Delta^2}{2 E_k^3}.
\end{align}
which gives
\begin{equation}
    \mathcal{J} = \left. \frac{dN}{d\lambda} \right|_{\Delta}  = \sum_k \frac{\Delta^2}{E_k^3}= 4 \sum{k} \frac{u_k^2 v_k^2}{E_k}.
\end{equation}
It relates the moment of the inertia to the pairing gap $\Delta$ and the quasiparticle energies $E_k$.

\subsection{Sign of the curvature and mid-shell symmetry}
A point of potential confusion arises regarding the sign of the moment of inertia $\mathcal{J}$ and the ``bending" of the energy parabolas in Fig. \ref{pair-rotate}. Equations (\ref{quad}) and (\ref{exact-energy}) both contain positive quadratic coefficients, yet they describe opposite curvatures in the energy plots: the ground-state energy $E(A)$ in the left panel is fitted as a parabola that "bends up" (concave-up). However, microscopic models like Equation (\ref{exact-energy})  and the BCS single-$j$ shell approximation predict a ``repulsive" quadratic term that suggests the energy curve should ``bend down". The bending direction (concavity) of the quadratic residual is determined solely by the sign of the second derivative (equivalently, the quadratic coefficient) and is invariant under translations of the coordinate origin or shifts of the reference value.

The symmetry of these parabolas is a direct consequence of particle-hole conjugation at the mid-shell. This symmetry governs the "counting" of valence particles/holes. A critical distinction lies in whether the system resists or encourages adding particles, which determines the sign of the pairing moment of inertia $\mathcal{J}_{pair}$.  For like-nucleon pairing, the quadratic ``repulsion" term, which arises from the reduction of available phase space due to the Pauli exclusion principle, ensures that the chemical potential $\lambda$ increases as particles are added ($d\lambda/dN > 0$), resulting in a positive stiffness. That accumulates to its maximum relative effect at the mid-shell. Here, the chemical potential $\lambda$ is exactly equal to the single-particle energy $\epsilon$ at the mid-shell $(N=\Omega)$ in a single-$j$ shell model. 

Conversely, for bosons with attractive interactions, the absence of Pauli blocking allows particles to condense into the same low-energy state. The quasi bosonic nature of $\alpha$ correlations makes it energetically favorable to add more particles, causing the chemical potential to decrease as particles are added ($d\lambda/dN < 0$). This leads to a negative moment of inertia. While a negative moment of inertia ($\mathcal{J} < 0$) implies unstable rotation in classical mechanics, in gauge space it physically distinguishes a Bose condensate from a Fermi liquid, enabling the coherent accumulation of energy. This, in turn, causes the quadratic “attraction” and the total correlation energy to reach a maximum at mid-shell, resulting in a smooth, nearly universal parabolic curvature that is the inverse of the like-nucleon pairing trend. While a negative moment of inertia ($\mathcal{J} < 0$) implies an unstable rotation in classical mechanics, in gauge space it physically distinguishes a Bose condensate from a Fermi liquid.

\bibliography{ref}

\end{document}